\documentclass{article}
\usepackage{graphicx} 
\usepackage{amsmath}
\usepackage[T1]{fontenc}
\usepackage{hyperref}
\usepackage{array}
\usepackage{booktabs}
\DeclareMathOperator*{\argmax}{arg\,max}

\title{Application Performance Benchmarks for Quantum Computers
}
\author{Krzysztof Kurowski \\  \href{mailto:krzysztof.kurowski@man.poznan.pl}{krzysztof.kurowski@man.poznan.pl} 
\and Piotr Rydlichowski \\  \href{mailto:prydlich@man.poznan.pl}{prydlich@man.poznan.pl} 
\and Konrad Wojciechowski \\ \href{mailto:kwojciechowski@man.poznan.pl}{kwojciechowski@man.poznan.pl} 
\and Tomasz Pecyna \\  \href{mailto:tpecyna@man.poznan.pl}{tpecyna@man.poznan.pl} 
\and Mateusz Slysz \\  \href{mailto:mslysz@man.poznan.pl}{mslysz@man.poznan.pl} 
}
\date{%
    Poznań Supercomputing and Networking Center, IBCH PAS\\Poznań, Poland\\%
}

\begin{document}

\maketitle

Current technological advancements in quantum computers highlight the need for application-driven, practical and well-defined performance benchmarking methods. As the existing NISQ device's quality of two-qubit gate errors rate is around 0.1\% -1\% and the number of qubits is still limited to a few or several dozen, naturally, we need to propose rather small algorithms instances taken from key promising application areas, such as quantum chemistry, combinatorial optimisation or machine learning. 
While many techniques for assessing the performance of logical components, such as gate fidelity and qubit coherence exist, it is still challenging to extrapolate those values onto the performance of different quantum algorithms and subroutines. This work aims to introduce a series of initial quantum application benchmarks together with a methodology of execution for measuring the performance and fidelity of the results. The proposed suite refers to several variational algorithms widely used on available NISQ devices but also includes examples of quantum circuits designed for a fault-tolerant quantum computer.


\section{Introduction}
Based on the recent developments in the physical implementations of quantum computers still in the NISQ era, we can observe the growing interest of different scientific communities and industries in designing accurate benchmarking routines. With each new quantum device or its upgraded version, it is certainly beneficial to access standardized performance evaluation methods, which allow for comparing the overall performance and tracking improvements across different quantum computer architectures. However, designing such benchmarks is challenging, especially given the vast dissimilarities in technologies employed by specific quantum platforms and the susceptibility of current machines to different noise sources.

This challenging problem has been addressed by some recent works, such as
\cite{lubinski2023applicationoriented, lubinski2023optimization, Proctor_2021}, while others have identified guidelines for creating valid benchmarks \cite{amico2023defining}. Nevertheless, this paper presents a new quantum benchmarking suite focused on application-oriented quantum circuits. The main objective is to collect and share quantum circuits with different properties and structures to measure quantum hardware's performance in various commonly used application scenarios. 

\subsection{Common and hardware-oriented quantum performance metrics}

Among the various metrics used to evaluate the quality and performance of quantum computers, two metrics of particular interest are Quantum Volume and Circuit Layer Operations per Second. These metrics hold significance in the context of quantum computing, and in the following discussion, we can't ignore their importance and implications for assessing the capabilities and efficiency of quantum devices.

\subsubsection{Quantum Volume}

Quantum Volume (QV) \cite{Cross_2019} is one of the most widely used metrics for estimating quantum computer capabilities. It captures the maximum size of a square quantum circuit that can be executed on a quantum computer with an output probability sufficiently similar to the output of the same circuit simulated classically. Quantum Volume $V_Q$ can be expressed with the following formula:

$$ \log_2 V_Q = {\argmax}_m \min (m, d(m)) $$

where $m$ is the width of the Haar random circuit and $d(m)$ is the depth. The execution of a QV circuit is successful when the heavy-output probability $h_U$ is greater than $2/3$, where heavy outputs $H_U$ have probabilities above the median value of the ideal distribution. This ideal distribution is generated by classically simulating the circuit with exponential overhead.

Even though many existing quantum algorithms typically do not consist of random circuits, the methodology behind QV benchmark assumes that such circuits can represent generic state preparation routines. Moreover, their structure, which utilizes two-qubit unitary gates, resembles circuits commonly used, such as variational methods and quantum adiabatic optimization \cite{Cerezo_2021}. These properties highlight QV's usefulness as a single-number metric for benchmarking quantum computers.

\subsubsection{CLOPS}

While QV is a holistic measure of a quantum computer's performance encompassing factors like capacity and quality of final results, it was also imperative to account for the speed at which these results were achieved. This is where CLOPS was introduced as the dedicated metric for quantifying the quantum computer's processing speed and computational efficiency \cite{wack2021scale}. 

CLOPS relies on layers, representing the basic blocks of parallel gate operations required to complete the quantum circuit. These layers are based on the QV layers, comprising the circuit depth. The formula for CLOPS is as follows:

\begin{equation}
    CLOPS = \frac{M \times K \times S \times D}{\text{time\_taken}},
\end{equation}

where $M$ denotes number of circuit templates, $K$ is the number of parameter updates, $S$ is a number of shots and $D$ is a number of QV layers.

Although the metric might appear straightforward, it captures various potential performance problems that a circuit may encounter, particularly in scenarios involving parametrized circuits and complex algorithms demanding efficient execution, making it a noteworthy metric for comparing quantum devices.

\section{Assumptions}

Adhering to the previously outlined metrics, we define circuit depth as the number of parallelly executed gates called layers that must be sequentially applied to compute a circuit. Following best practices presented in \cite{lubinski2023applicationoriented}, we take the {Rx, Ry, Rz, CNOT} set as a basis gate set. We also assume that the CNOT gates can be applied between any pair of qubits. 

We opted to assess quantum computers primarily based on specific application-oriented use cases. We will employ the fidelity measure to minimize the risk of benchmarking algorithms rather than the quantum computer's true capabilities. This measure gauges the quantum computer's ability to execute the described circuits by comparing the resulting measurement distribution $P_\text{output}$ to the ideal one $P_\text{ideal}$  obtained from an exact state vector simulation.

For the purpose of selected benchmarks, the average fidelity is defined as:

\begin{equation}
    F(P_\text{ideal}, P_\text{output}) = \max\big\{ F_\text{raw}(P_\text{ideal}, P_\text{output}), 0\big\}
\end{equation}

where $F_\text{raw}$ is defined to punish distributions which are close to uniform:

\begin{equation}
    F_\text{raw}(P_\text{ideal}, P_\text{output}) = \frac{F_s(P_\text{ideal}, P_\text{output}) - F_s(P_\text{ideal}, P_\text{uni})}{1 - F_s(P_\text{ideal}, P_\text{uni})}
\end{equation}

and the $F_s$, defined below, is the standard measure of classical fidelity of two probability distributions (related to the Hellinger distance):

\begin{equation}
    F_\text{s}(P_\text{ideal}, P_\text{output}) = \left( \sum_x \sqrt{P_\text{output}(x)P_\text{ideal}(x)} \right)^2
\end{equation}

where $P_\text{output}(x)$ and $P_\text{ideal}(x)$  are the respective probabilities of observing bit string $x$.

Using the classical fidelity $F_s$ has a serious drawback in that it can yield non-zero values for random (uniform) distributions. On the other hand, we can see that normalizing makes $F(P_\text{ideal}, P_\text{uni}) = 0$, which is useful for assessing errors in quantum hardware, especially as circuits become larger or more complex. In such cases of significant decoherence, the output distribution approaches the uniform and, therefore, should be punished. At the same time, it is worth noting that because of the way $F_\text{raw}$ is formulated, benchmarks for which $F_s(P_\text{ideal}, P_\text{uni}) \approx 1$ should be avoided. This is not an issue with the methodology, as such benchmarks are not useful for assessing the quantum properties of any system.

It has also been observed that the normalized fidelities of different benchmarks with similar circuit shapes show a higher correlation than the standard fidelities. Therefore, using normalized fidelity is more practical and informative for evaluating quantum computing results.

\section{Execution}

We identify a set of quantum algorithms for the application benchmarks, exemplifying the common approaches to performing quantum computations in different application areas. These include the currently most commonly used near-term variational algorithms and routines in fault-tolerant quantum computing.

Since the benchmarking suite includes hybrid and purely quantum algorithms, we must agree on a base methodology applied in each case. While this methodology is designed to be as general as possible, there are still cases where such an approach cannot capture the whole nature of the challenge posed to the quantum machine. It will be elaborated on in subsequent sections, where applicable. 

To evaluate a given quantum system’s ability to execute an algorithm, we choose a single quantum circuit, which by design is meant to represent the most typical single execution, either hybrid or purely quantum. It is especially noteworthy in the case of variational algorithms, where we do not intend to perform a full run, optimizing the parameters, but rather fix the parameters in place and estimate the fidelity on a single non-parameterized circuit. It is done carefully to avoid cases where the ideal distribution is close to uniform.

The logical quantum circuits for each application benchmark are compiled into OpenQASM 2.0/3.0 assuming all-to-all connectivity and {Rx, Ry, Rz, CNOT} as the basis gate set. For execution on real quantum backend devices, these circuits can be freely recompiled and optimized as long as they remain logically equivalent to the ones delivered within the described suite. For this assumption to be valid, we choose circuits that represent unitary matrices sufficiently different from the identity. Circuit approximation techniques are allowed as a way of finding more efficient compilations, but in the end, the same fidelity measures apply. While error mitigation is not meant to be a part of these benchmarks, error suppression techniques like dynamic decoupling can be used.

Each circuit representing a specific application benchmark is meant to be measured at least 1000 times, so the appropriate measurement average for estimating specified metrics has to be taken. No error mitigation has to be applied.

We assume success for each of the following benchmarks when the fidelity surpasses a specified threshold, which can differ in different cases. These criteria are set to ensure a minimum level of complexity and capability in the quantum computations being performed. By selecting these thresholds, we aim to evaluate and compare the performance of quantum systems that can handle moderately sized circuits and exhibit a reasonable level of fidelity in their results. These benchmarks provide a standardized measure to assess the progress and advancements in quantum computing technology. 

The following main metrics based on best practices discussed in \cite{lubinski2023applicationoriented} have been identified:

\begin{itemize}
    \item \textbf{Execution Time:} time spent on a quantum simulator or hardware backend running the circuit;
    \item \textbf{Circuit Depth:} depth of the circuit after transpiling it to the basis gates set defined as {Rx, Ry, Rz, CNOT}
    \item \textbf{Fidelity:} a measure of how well the simulator or hardware runs a particular benchmark;
\end{itemize}

For all the following benchmarks, we share the corresponding quantum circuits in both openQASM 2.0 and openQASM 3.0 formats. The codes are available online at \url{https://drive.man.poznan.pl/f/6394136}.




\subsection{Entanglement in GHZ state}

Entanglement is a key property differentiating quantum systems from purely classical ones. It is known that quantum systems containing sufficiently low amount of entanglement can be simulated efficiently on a classical computer. Because of that, the ability of a quantum computer to generate genuine multipartite entanglement is essential for them to be able to outperform their classical counterparts.

To this end, similarly to \cite{moses2023race}, we propose producing Greenberger-Horne-Zeilinger (GHZ) states as a benchmark of the quantum system’s ability to entangle multiple qubits. These states have a convenient property that they can exhibit genuine multipartite entanglement while, at the same time, their fidelity can be efficiently estimated. While in other benchmarks, we used a measure of fidelity based on the Hellinger distance, in this case, it is not sufficient to detect genuine entanglement. Therefore, we employ instead a standard approach used in other such experiments performed on various quantum devices \cite{mooney2021generation, monz201114}

The N-qubit GHZ state is defined as:
\begin{equation}
|\text{GHZ}_N \rangle = \frac{1}{\sqrt{2}}\left(|0\rangle^{\otimes N} + |1\rangle^{\otimes N}\right),    
\end{equation}

where $N$ is the number of qubits.

The fidelity F of this state can be expressed with:

\begin{equation}
    F = \frac{1}{2}(P + C)
\end{equation}
,
where the \textit{population} $P = \rho_{00\dots 0,00\dots 0} + \rho_{11\dots 1,11\dots 1}$ is measured as the sum of occurrences of outcomes $|00\dots 0\rangle$ and $|11\dots 1\rangle$, while \textit{coherence} $C = |\rho_{00\dots 0,00\dots 0}| + |\rho_{11\dots 1,11\dots 1}|$ can be estimated either through Multiple Quantum Coherences (MQC) or parity oscillations \cite{mooney2021generation}.

While performing the benchmark, we do not enforce a definite circuit construction or a method for calculating the coherence C. Since this test aims to examine the quantum system’s ability to create entangled states, the specific techniques can be tailored to a specific device to achieve the highest possible fidelities.

\begin{figure}
    \centering
    \includegraphics[width=0.5\linewidth]{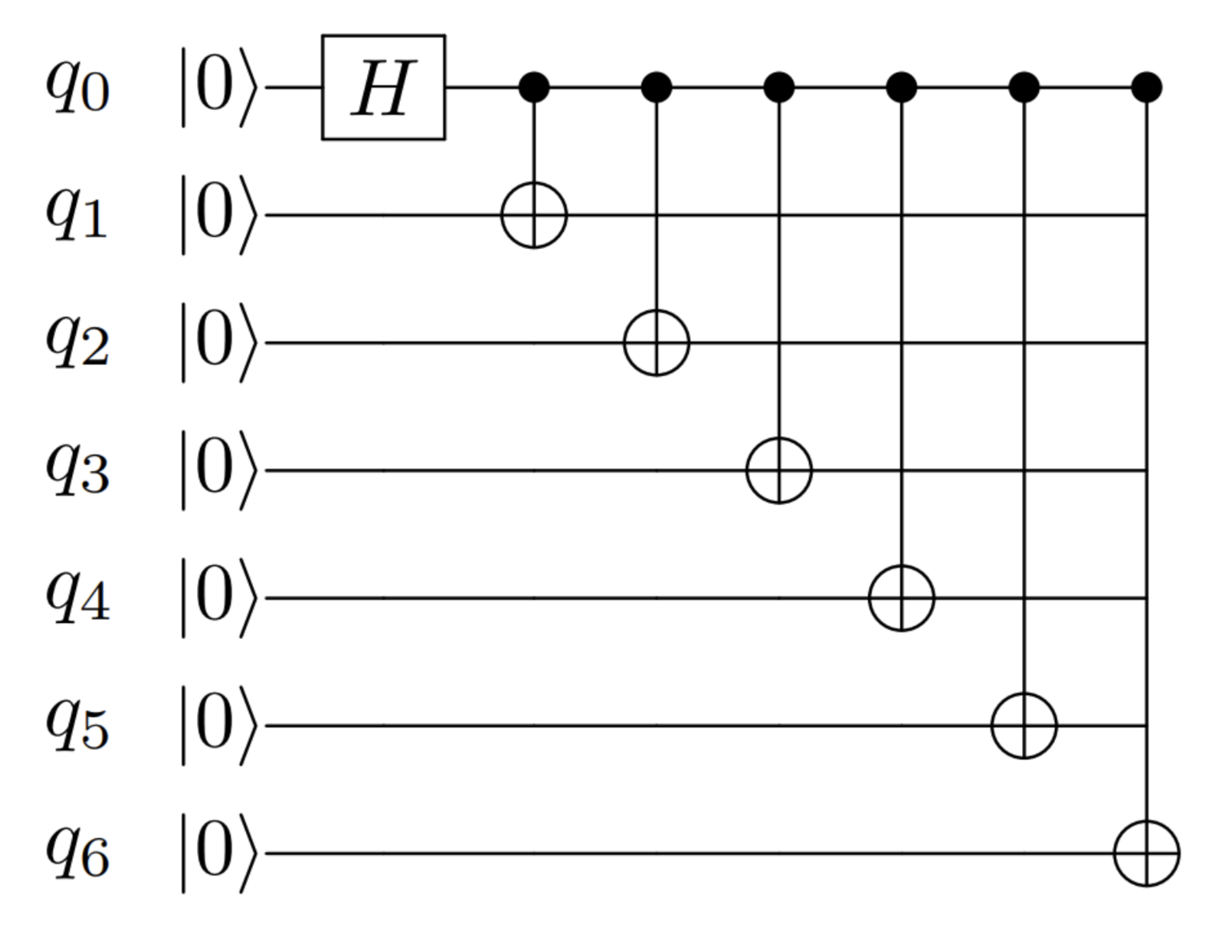}
    \caption{The example 7-qubits GHZ state circuit.}
    \label{fig:GHZ_circuit}
\end{figure}

\subsection{Toffoli gate}

The $n$-qubit Toffoli gate is a multi-qubit quantum gate operation that acts on $n$ qubits, with $n-1$ control qubits and $1$ target qubit. The action of the Toffoli gate is simply that the target qubit is inverted if all control qubits are in the state $1$. Otherwise, an identity operation is performed. The $n$-qubit Toffoli gate is an essential multi-qubit operation that can be used in important applications such as Grover search algorithm \cite{grover1996fast}, Quantum Fourier Transformation \cite{nam2020approximate}, Shor’s number-factoring algorithm \cite{van2005fast} and quantum error correction \cite{cory1998experimental}. The decomposition into basic 1-qubit and 2-qubit gates scales quadratically ($n^2$) in the number of required $2$-qubit gates \cite{shende2008cnot}. The quadratic scaling puts hard fidelity requirements on the system. Although there are techniques like auxiliary qubits assisted implementations \cite{he2017decompositions}, that lead to linear ($n$) overhead of $2$-qubit gates, or native implementation in trapped ion system \cite{fang2023realization}, here we only focus on the performance of an $n$-qubit Toffoli gate, independent of its implementation.
\\
\\
\textit{Problem instance:}

A rigorous characterization of an $n$-qubit Toffoli gate can be done using quantum process tomography, which is highly inefficient ($12^n$ measurements required) and cannot be implemented on current devices. We therefore propose to measure only the truth-table (similar as in \cite{fang2023realization}), which is also not efficient ($2^n$ measurements required), but doable for NISQ devices. The average success probability $F$ (average over all possible input states) to measure the correct output state should be $F>0.5$. The problem instance should implement a $6$-qubit Toffoli gate with circuit approximation and $5$-qubit Toffoli gate without circuit approximation.

\subsection{Grover’s Algorithm}

Grover’s algorithm \cite{grover1996fast} remains one of the most well-known quantum algorithms. It solves the problem of unstructured search using quadratically fewer calls to the oracle. In classical computation, $O(N)$ evaluations of a black box function are required, while the quantum method needs only $O(\sqrt{N})$. It has also been found that this algorithm is asymptotically optimal.
\\
\\
\textit{Problem instance:}

The circuit for Grover’s algorithm generally consists of two key parts: the quantum oracle $U_\omega$, which marks the solution states, and the diffusion operator, which allows for manipulating the qubits to increase the amplitude of the marked state. In the proposed benchmark, we employ a simple $3$-qubit circuit for finding bitstrings marked by the oracle. Compared to the latter ones described in this document, the novelty introduced by this circuit are three-qubit gates, namely CCZ gate, which can be decomposed into a Toffoli gate and two Hadamard gates. This supplements the proposed benchmark suite with the possibility of testing performance and compilation effectiveness for quantum circuits where such gates are essential.

\begin{figure}
    \centering
    \includegraphics[width=0.5\linewidth]{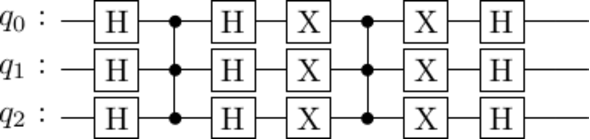}
    \caption{The example 3-qubit Grover circuit.}
    \label{fig:grover_alg}
\end{figure}

\subsection{Quantum Fourier Transform}

The Quantum Fourier Transform (QFT) \cite{coppersmith2002approximate} is a fundamental quantum algorithm used for performing a Fourier transformation on a quantum state. It is a key component in many quantum algorithms, particularly in quantum algorithms for prime factorization, quantum phase estimation, and quantum simulation. The QFT maps an input state $|x\rangle$ to its Fourier-transformed state $|y\rangle$, where $y = F(x)$ and $F$ denotes the Quantum Fourier Transform operator. 

The QFT algorithm operates on a register of n qubits, and the input state is typically encoded in the amplitudes of the computational basis states. The QFT applies a sequence of controlled rotations and Hadamard gates to transform the input state into its Fourier-transformed state. The QFT can be represented as a unitary matrix, which depends on the number of qubits in the register. 

By applying the QFT to an input state, one can extract information about the frequency components of the state. This is particularly useful in applications such as signal processing, data compression, and solving certain mathematical problems efficiently \cite{nam2020approximate}. The performance of the QFT is influenced by factors such as the number of qubits, gate errors, and coherence times of the quantum system. Achieving high fidelity and minimizing errors are crucial for obtaining accurate results from the QFT.
\\
\\
\textit{Problem instance:}

In the QFT benchmark, we specifically focus on running the inverse QFT. The inverse QFT is applied to a quantum state that is initially prepared in a Fourier basis state. However, rather than utilizing the QFT to create this state, we employ a series of one-qubit gates, such as Hadamard gates and Z rotations, to encode a specific integer value $x$ in the Fourier basis. This approach allows us to evaluate the performance of the inverse QFT circuit in accurately decoding the encoded integer value.

\subsection{VQE for quantum chemistry calculations}

The Variational Quantum Eigensolver (VQE) is a quantum algorithm designed to solve problems in many domains, including but not limited to quantum chemistry, quantum simulations or optimization. It is an example of a hybrid quantum-classical algorithm that combines quantum computing with classical optimization techniques \cite{tilly2022variational}.

In many VQE applications, the primary goal is to find a given molecular system's ground state energy and corresponding wave function. This is a crucial task in quantum chemistry as it provides insights into molecules' electronic structure and properties, which are vital for various applications such as drug discovery, materials science, and catalysis.

One of the most significant advantages of VQE is its potential to leverage near-term quantum devices, even with limited qubit counts, connectivity, and high error rates. VQE-based approaches are part of an active area of research and development, with ongoing efforts focused on improving their scalability, robustness against noise and errors, and enhancing their applications in various domains. As quantum hardware continues to advance, VQE holds the promise of being able to solve complex quantum mechanical problems on the NISQ-era quantum devices.

In the proposed application benchmark, the quantum device executes circuits typically used for estimating the energy of a molecule. The results are scored based on the average fidelity metric calculated from the measurements. This allows evaluating the performance of VQE on the quantum machine while setting aside the difficulties related to measurement scaling in the energy estimation process, which is less favourable than the fault-tolerant Quantum Phase Estimation (QPE) algorithm.

The electronic Hamiltonian of a molecular system, before it is used in VQE, is most commonly written in the second quantized form, using the fermionic creation and annihilation operators:

\begin{equation}
    H_{el} = \sum_{p, q}h_{pq}a_p^\dagger a_q+ \frac{1}{2}\sum_{p,q,r,s}h_{pqrs}a_p^\dagger a_q^\dagger a_r a_s
\end{equation}

The first term in this equation corresponds to single-electron excitations, and the second term corresponds to two-electron excitations. Coefficients $h_{pq}$ and $h_{pqrs}$ are one- and two-electron integrals.
This Hamiltonian is then mapped to qubits using well-known transformations, resulting in an operator written as a sum of products of Pauli matrices denoted as $\sigma$

\begin{equation}
    H = \sum_j \alpha_j P_j = \sum_j \alpha_j \prod_i \sigma_i^j
\end{equation}
\\
\\
\textit{Problem instance:}

The Unitary Coupled Cluster with Singles and Doubles is defined as:

\begin{equation}
    | \phi(\theta)\rangle = e^{T - T^\dagger}|\phi_{HF}\rangle
 \end{equation}

where $T(\theta)$ is the cluster operator, $|\phi \rangle_{HF}$ is the reference state, chosen here to be the Hartree-Fock state. 

The cluster operator $T(\theta)$ has the following definition:
\begin{equation}
    T(\vec{\theta}) = T_1(\vec{\theta_1}) + T_2(\vec{\theta_2}),
\end{equation}

where

\begin{equation}
    T_1(\vec{\theta_1}) = \sum_{i, j}\theta_{ij}a_j^\dagger a_j
\end{equation}

\begin{equation}
    T_2(\vec{\theta_2}) = \sum_{i, j, k, l}\theta_{ijkl}a_i^\dagger a_j^\dagger a_k a_l
\end{equation}
	
Although this type of ansatz has unfavourable scaling of the required number of gates with an increasing number of electrons and spin orbitals, it is often used for small-scale demonstrations of VQE and also serves as a basis for more efficient approaches, e.g. AdaptVQE, which allows to decrease the depth of the circuit significantly.

We evaluate the fidelity of the quantum states prepared on a quantum computer with a selected ansatz for LiH molecule in the minimal STO-3G basis set. Active space reduction is also used to execute this benchmark on three qubits for additional flexibility.

\subsection{QAOA for combinatorial problems optimization}

\begin{figure}
    \centering
    \includegraphics[width=0.5\linewidth]{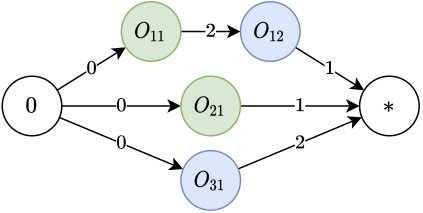}
    \caption{The small JSSP problem instance is represented as a disjunctive graph (colours represent different machines).}
    \label{fig:qaoa_toy_instance}
\end{figure}

The Quantum Approximate Optimization Algorithm (QAOA) is a quantum algorithm designed to solve combinatorial optimization problems. It is based on the framework of variational quantum algorithms and is often used in the context of quantum computing. QAOA combines classical optimization techniques with quantum operations to find approximate solutions to optimization problems. 

The algorithm begins with an initial state prepared as a superposition of computational basis states. This state is typically denoted as $|+\rangle^{\otimes n}$, where n is the number of qubits. QAOA consists of a sequence of p layers, where each layer consists of two types of quantum Hamiltonians: the problem Hamiltonian $H_P$ and the mixing Hamiltonian $H_M$. The problem Hamiltonian represents the objective function of the optimization problem, while the mixing Hamiltonian helps explore different solutions. 

The problem Hamiltonian is applied to the state at each layer, followed by the mixing Hamiltonian. The evolution of the state is controlled by parameters $\gamma$ and $\beta$, which are optimized to minimize the objective function. The state is measured after $p$ layers, and the measurement outcomes are used to approximate the optimal solution. 

The output of the QAOA algorithm is an approximate solution to the optimization problem. The quality of the solution depends on the number of layers and the parameters $\gamma$ and $\beta$. By increasing the number of layers and using classical optimization techniques to refine the parameters, QAOA can provide increasingly better approximations to the optimal solution.

The job shop scheduling problem is a classic combinatorial optimization problem that involves determining the optimal sequence of operations for a set of jobs to be processed on a set of machines. Each job consists of multiple operations that require specific processing times on different machines. The objective is to minimize the overall makespan, which is the total time required to complete all jobs. Job shop scheduling problems are known for their complexity due to the presence of constraints such as machine availability, precedence relationships between operations, and resource limitations. Efficiently solving job shop scheduling problems has practical applications in various industries, such as manufacturing, logistics, and project management.
\\
\\
\textit{Problem instance:}

The problem that will be considered as a benchmark will be as following:
\begin{itemize}
    \item Job $1$:
    \begin{itemize}
        \item Operation $1$, Time: $1$ unit, Machine $1$
        \item Operation $2$, Time: $2$ units, Machine $2$
    \end{itemize}
    \item Job $2$:
    \begin{itemize}
        \item Operation $1$, Time $1$, Machine $1$
    \end{itemize}
    \item Job 3:
    \begin{itemize}
        \item Operation $1$, Time $2$, Machine $2$
    \end{itemize}
\end{itemize}

This instance has the lowest completion time equal $3$. Setting maximum feasible time to $3$, it requires $7$ qubits to encode in the time-indexed representation described in \cite{kurowski2023application}. The depth of the QAOA circuit with p=1, which solves this circuit, is equal to $24$.

\subsection{QSVM for image classification}

Quantum Support Vector Machine (QSVM) is an algorithm that utilizes a quantum kernel for a Support Vector Machine (SVM) machine learning model. SVM is an algorithm for data classification and regression that finds an optimal hyperplane between different classes in a dataset. However, if the data is not linearly separable, a kernel function that maps the dataset to a higher dimensional feature space is needed for the SVM algorithm to perform correctly. 

The quantum part of the algorithm finds a feature map using a quantum computer, which can later be used to create a kernel matrix. In theory, a quantum feature map can extract complicated patterns from data that would not be possible by using only classical transformations for a kernel function. The quantum circuit used in this algorithm can be divided into several components. It consists of 2 symmetric parts - each corresponding to one of the elements of the pair of variables for which the value of the kernel function is calculated. Each part consists of a block that encodes data using X and Z rotation gates, a block that generates quantum entanglement, and a block of parameterized Y rotation gates with trainable parameters. As the encoding of variables can be done densely by putting $2$ numerical variables per qubit, the number of qubits for a given instance size equals the number of variables divided by $2$. A crucial parameter for a quantum computer is the number of qubits, enabling it to fit larger data onto the quantum computer.

The depth of quantum circuits used in this algorithm is similar between instance sizes and is around $10$. The entanglement strategy can influence it a little, to a factor of up to $20$. The full form of the circuit consists of a parameterized block followed by its inverse and after assigning the parameters based on the input data, the resulting circuit can be represented by a unitary matrix close to the identity. To avoid this, we use only the first part of the circuit, to test the quantum machine’s ability to create such entangled states.
\\
\\
\textit{Problem instance:}

In the proposed application, the benchmark QSVM algorithm is applied to a well-known benchmark dataset for image classification - MNIST \cite{slysz2023exploring}. It is a collection of grayscale images of handwritten digits, with resolution $28\times28$ pixels. The images are downscaled to a smaller resolution to fit the data onto the quantum computer. The binary optimisation problem is solved with the minimum accuracy for the smallest image size ($4\times4$).should be able to achieve more than $70\%$. To fit $16$ variables, the minimum number of qubits is $8$.

\begin{figure}
    \centering
    \includegraphics[width=0.5\linewidth]{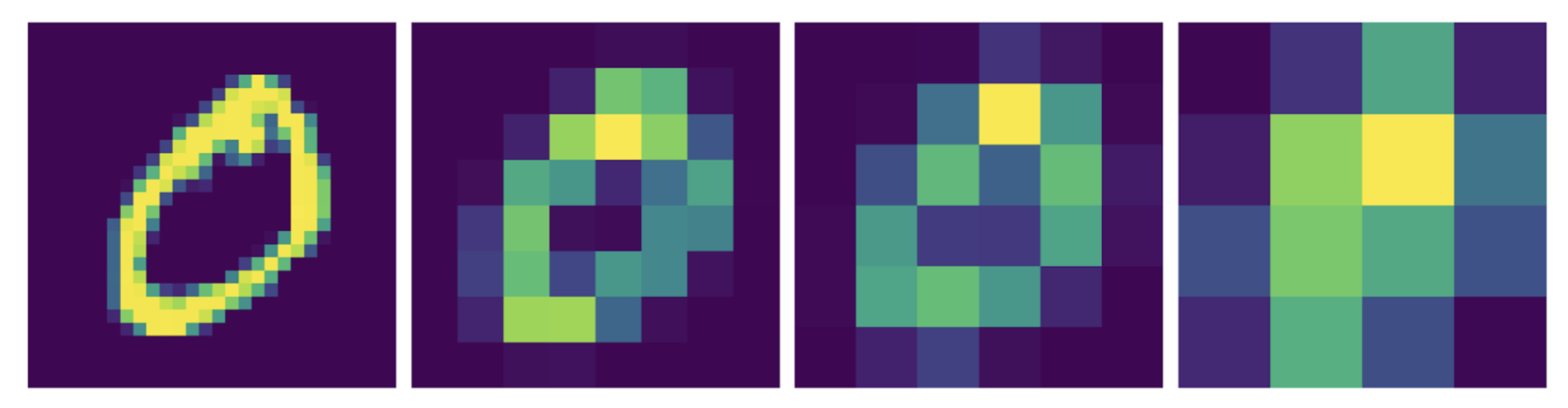}
    \caption{A grid of original size and resized images of a digit '0' in different resolutions.}
    \label{fig:mnist0}
\end{figure}

%

\section{Conclusion}

This paper emphasizes the need for standardized application performance benchmarks for quantum computers. It acknowledges the challenges of assessing quantum computer performance due to technological variations and noise susceptibility. The document introduces a series of quantum application benchmarks and a methodology for measuring performance and fidelity. The proposed benchmarks encompass a variety of quantum algorithms, from variational algorithms used in near-term quantum devices to circuits designed for fault-tolerant quantum computers. Each benchmark's problem instance is defined, showcasing the diversity of quantum applications.

In conclusion, these standardized benchmarks are essential for evaluating and comparing quantum computing technologies, tracking their progress, and enabling applications in various domains. They contribute to developing a benchmarking framework that ensures the reliability and effectiveness of quantum computing solutions. 

Finally, as one of the targets of the EuroHPC JU is to develop and support a highly competitive and innovative quantum computing ecosystem broadly distributed in Europe capable of autonomously producing quantum computing technologies and architectures and their integration on leading HPC computing systems, the proposed application performance benchmarking suite together with the open source code are also published to evaluate upcoming quantum systems installations, in particular a trapped ions quantum system hosted at the Poznan Supercomputing and Networking Center (PSNC), in Poland, within the EuroQCS-Poland consortium.

\bibliographystyle{unsrt} 
\bibliography{refs}

\end{document}